# Implicit Phonon Shifts and Thermodynamical Properties of Rigid Carbon Nanotube Ropes

**Shuchi Gupta, K. Dharamvir and V. K. Jindal***

Department of Physics, Panjab University, Chandigarh – 160014

*We calculate phonon shifts of external modes of a bunch of carbon nanotubes. A simple model based on atom-atom potential has been used to calculate the implicit anharmonicity in the phonons of carbon nanotube bundles having rigid tubes, with the assumption that under hydrostatic pressure only the intertube distance in the bunch varies. Such a model is important as long carbon nanotube ropes will be an extension of a fixed length ropes as is done here. Various bulk and thermodynamic properties like thermal expansion, bulk modulus and the Gruneisen constants and external phonon shifts which naturally enter into the calculation are also described and compared with the available data. The specific heat capacity has also been calculated.*

## 1. Introduction

The large variety of carbon allotropes, as well as their present potential applications in diverse fields such as nanoelectronics[1] and bioengineering[2], gives them a special place among all the elements. Since their discovery in 1991[3], carbon nanotubes (CNT) have hogged much attention from various research communities. Their high aspect ratio, large tensile strength, the ability of their existence in either metallic or semiconducting forms and extreme flexibility make them promising candidates as high strength fibers and various novel nanometer scale electronic and mechanical devices.

---

* Author with whom correspondence be made, e-mail: jindal@pu.ac.in. Present temporary address: Institute for Chemistry, Technical University, Berlin, D-10623 Germany.

A CNT can be envisaged as a graphene sheet rolled up into a seamless cylinder to form a macromolecule. They exist as single nanotubes of various kinds as well as materials of these in forms of bunches of single wall (SWNT) or multiwall (MWNT) formations. In a bundle consisting of identical nanotubes, they are arranged in a two-dimensional hexagonal close packing (forming 'nanoropes'), as shown in fig. 1, interacting via weak Van der Waals type attractive forces [4].

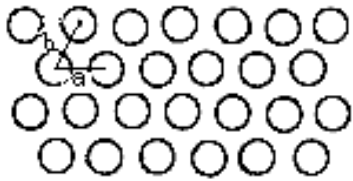

**Fig.1**: Cross-section through a bundle of carbon nanotube.

The variation of temperature and pressure results in changes in the ordered environment of the system which can undergo phase transitions. In addition, the increase in temperature and pressure exploits the anharmonicity of the potential, which modifies some of the observed low temperature effects and introduces new effects like thermal expansion and phonon shifts and widths [5]. Pressure and temperature dependence of structural and dynamical properties are very useful for interpreting and understanding theoretical models for the system under study. Various pressure effects on the isolated as well as on bundles of rigid and non-rigid tubes have been already reported in our earlier paper [6].

In carbon nanotubes, phonon dispersions provide most of the information that is needed to calculate the harmonic thermo dynamical quantities as well as the dominant contribution to thermal expansion and specific heat. To our knowledge, not much work has been done on the bunches, though Sauvajol et al[7] have attempted a calculation on the phonons in single wall carbon nanotube bundles. In this paper, considering only the weak Van der Waal (VdW) interactions between the atoms on different nanotubes in a bundle and neglecting the intramolecular interactions, various phonon related properties have been studied. We have calculated the phonon frequencies, Gruneisen parameters, thermal expansion, heat capacity of a variety of carbon nanotubes. The implicit phonon shift, which arises due to the implicit temperature dependence of the phonon frequencies and is obtained from the volume dependence of the phonon frequencies, have also been calculated.

## 2. Theoretical Model

We study the properties of a nanotube rope in which all the tubes are identical. Theoretical models used by us to explain the structure, dynamics and related properties of $C_{60}$, $C_{70}$ and nanotube material solids are based on model potentials for C-C interaction which are then summed over to obtain molecule-molecule potentials. There is no evidence for covalent bonding between molecules. This model also reproduces reasonably well, the observed structure, bulk and lattice properties of $C_{60}$ and $C_{70}$ solids [8]. In Ref. [6], we have already studied the structural transformations of CNT ropes on application of hydrostatic pressure. For temperature dependent studies of the dynamics and statics of CNT ropes, we use similar procedures as described in Ref. [6].

In a rope, CNTs arrange themselves in a 2-D hexagonal arrangement, as shown in fig. 1, (each tube surrounded by six other tubes along x and y directions) but each tube has some length also, which extends in z-direction, making it a 3D system. For our calculations, nanotubes of specific length have been taken and that length has been kept fixed since the nanotubes are considered to be rigid.

Our crystal of CNT is a variable area but fixed length 3-D system. Change in volume, therefore, corresponds to the change in area of the crystal. We can, therefore, apply 3-D formalism to get 2-D results.

## 2.1 Model Potential

The inter-tube potential energy $U_{l\kappa,l'\kappa'}$ between two nanotubes (molecules), identified by $\kappa$ molecule in unit cell index **l**, $\kappa'$ molecule in cell **l'**, can be written as a pair-wise sum of C-atom-atom potentials (C-C) on these two molecules, i.e.

$$U_{l\kappa,l'\kappa'} = \sum_{ij} V(r_{ij}), \qquad (1)$$

where the sum in Eq. (1) includes all the M atoms in each of the nanotube molecules, and V(r) is the C-C potential. Since ours is a monomolecular unit cell and therefore, here, $\kappa = 1$ and $\kappa' = 1$. We take the potential V(r), where r is the distance between the C-C atoms, to be given by:

$$V(r) = -A/r^6 + B\exp(-\alpha r) \qquad (2)$$

The interaction parameters A, B and $\alpha$ have been obtained for various atom-atom interactions from gas phase data, and we obtain these for our use from the set provided by Kitaigorodski [8]. For C-C interaction, these parameters have been tabulated in Table I.

**Table I:** Atom-atom potential parameters (Kitaigorodski [9])

| A=358 kcal/mole-A$^6$ | B=42000 kcal/mole | $\alpha$=3.58A$^{-1}$ |
|---|---|---|

The total potential energy $\Phi$ can be obtained by carrying out the lattice sum, knowing the position of the lattice points,

$$\Phi = 1/2 \sum_{l,l'}{}' U_{l\kappa,l'\kappa'} \qquad (3)$$

Knowing molecular atomic coordinates and crystal structure (here, it is a two-dimensional crystal with hexagonal symmetry), it is easy to compute the potential energy and structure that provides the minimum energy configuration.

Using this model potential, the harmonic phonon frequencies $\omega_{qj}$ (which under rigid molecule approximation, lead to only external phonons, hereinafter referred to as simply phonons) for values of q in the Brillouin zone can be obtained. It needs to be stressed here that the modes we now talk about are only e.g. in two surface dimensions, including the librational and translational modes.

An application of a hydrostatic pressure $p$ leads to a new volume (which in this case of fixed length of nanotubes, results in new area, and new volume results from this change in area only) which is obtained by minimizing the potential energy function

$$\Phi_p = \Phi(V) + p\Delta V \qquad (4)$$

This structure obtained after minimising the potential energy gives the new volume and the dynamical matrix is again diagonalized corresponding to this structure. The pressure dependent phonon frequencies are thus obtained which correspond to volume dependent phonon frequencies. From the changes in the phonon frequency, the Gruneisen parameters, $\gamma_{qj}$, can be easily calculated, defined as

$$\gamma_{qj} = -\frac{\partial \ln \omega_{qj}}{\partial \ln V} \qquad (5)$$

If one takes a single nanotube as one molecule in the unit cell, the energy of the solid will depend on how long nanotubes are taken. It is therefore appropriate to define energy per atom, and use mass of a single carbon atom to calculate frequency of the modes, which in case will be frequency of vibrational and librational modes of the mono-molecule unit cell with motion limited to surface (a 2-D motion). The thermodynamic properties based on this dynamics are easily calculable from the Helmholtz free energy function. Its quasi – harmonic part is given by

$$F_{qh} = kT \sum_{qj} \ln\left(2\sinh\frac{\hbar\omega_{qj}}{kT}\right) + \Phi(V) \qquad (6)$$

where $\Phi(V)$ is the static lattice contribution. The sum over $q$ runs over first Brillouin zone and $j$ sums over the phonon branches, defined for a hexagonal surface.

**2.2 Harmonic Phonons**

The nanotube material is a molecular crystal, having vibrational and librational modes. The total potential energy of the crystal is dependent upon the intermolecular separation as well as on the orientation of the molecules. Since each molecule in 2-D unit cell has its length extending in the third direction and therefore for the sake of convenience and generality with 3-D systems, we describe its lattice dynamics also by a Taylor series expansion in terms of a six-component translation-rotation displacement vector $u_\mu(l\kappa)$ (instead of that suitable for a 2-D system) representing $\mu$ th component of the displacement of $\kappa$ th molecule in **l** th cell. Here the index $\mu$ runs from 1 to 6, the first three components representing the translational displacement (the x, y and z components) and the other three, the rotational displacements as angles about the three axes. The three axes are profitably chosen as the principal axes of moment of inertia of a molecule, as this simplifies the expressions for the crystal Hamiltonian. The expression for the molecular crystal Hamiltonian is then written as [5]

$$H = \frac{1}{2}\sum_{l\kappa\mu} m_\mu(\kappa)\left[\dot{u}_\mu(l\kappa)\right]^2 + \frac{1}{2}\sum_{l_1 l_2}\sum_{\kappa_1 \kappa_2}\sum_{\mu_1 \mu_2} \Phi_{\mu_1\mu_2}(l_1\kappa_1, l_2\kappa_2) u_{\mu_1}(l_1\kappa_2) u_{\mu_2}(l_2\kappa_2), \qquad (7)$$

where we retain only the harmonic part of the potential energy. The kinetic energy part involves the translational energy for $\mu \leq 3$, when $m_\mu(\kappa)$ represents the mass of the molecule at $\kappa$, which in present case is independent of $\kappa$. For $\mu > 3$, $m_\mu(\kappa)$ represents the moment of inertia along the principal axes for the $\kappa$ th molecule and kinetic energy corresponds to rotational kinetic energy. $\Phi_{\mu_1\mu_2}(l_1\kappa, l_2\kappa_2)$ is a harmonic force constant, and is defined in terms of second derivative of the potential energy at equilibrium, i.e.

$$\Phi_{\mu_1\mu_2}(l_1\kappa, l_2\kappa_2)_1 = \left(\frac{\partial^2 \Phi}{\partial u_{\mu_1}(l_1\kappa_1) \partial_{\mu_2}(l_2\kappa_2)}\right)_0 \qquad (8)$$

Some of the translation-rotation components especially translation along z-direction are redundant for our 2-D system.

Since $\Phi$ is related to atom-atom potential V(r), (Eqs.1-3), the force constant in Eq. (7) can be expressed in terms of V(r), with appropriate transformation relations involving the relationship of the atom-atom distance r as functions of molecule translation or rotations. Finally, the dynamical matrix defined as,

$$M_{\mu_1\mu_2}(\kappa, \kappa', \mathbf{q}) = \frac{1}{(m_{\mu_1}(\kappa) m_{\mu_2}(\kappa'))^{1/2}} \sum_{l'} \Phi_{\mu_1\mu_2}(l\kappa, l'\kappa') \exp[i\mathbf{q}.(\mathbf{R}(l'\kappa') - \mathbf{R}(l\kappa))] \qquad (9)$$

leads to the calculation of phonon frequencies $\omega_{\mathbf{q}j}$ and eigenvectors $\mathbf{e}(\kappa|\mathbf{q}j)$ for values of wave vector $\mathbf{q}$ in the Brillouin zone by diagonalisation of the above dynamical matrix.

## 2.3 Implicit anharmonic effects

Increase in temperature results in (a) increased amplitudes of vibrating units, and (b) increase in volume due to thermal expansion. The former is an explicit anharmonic effect, and its calculation requires inclusion of anharmonic terms in the Taylor series expansion of the potential energy. This results in modification of harmonic phonons through shifts and widths. The latter effect is called implicit anharmonic effect, and is normally dominant [5] and easy to calculate.

The implicit shift in the phonon frequencies arises due to the thermal expansion and is obtained from the volume dependence of the phonon frequencies. The implicit temperature dependence of the phonon frequencies is therefore written as

$$\left(\frac{\partial \omega_{qj}}{\partial T}\right)_{im} = \frac{\partial \omega_{qj}}{\partial V}\frac{\partial V}{\partial T} \qquad (10)$$

which when expressed in terms of the Gruneisen constants $\gamma_{qj}$ becomes

$$\left(\frac{\partial \omega_{qj}}{\partial T}\right)_{im} = -\gamma_{qj}\, \omega_{qj}\beta \qquad (11)$$

where $\beta$ is the volume thermal expansion coefficient.

$$\beta = \frac{1}{V}\frac{\partial V}{\partial T} \qquad (12)$$

Rewriting eq. (11) and integrating from T=0 to T, we obtain the expression for the fractional implicit phonon frequency shift as

$$\left(\frac{\Delta \omega_{qj}}{\omega_{qj}}\right)_{im} = \exp[-\gamma_{qj}\varepsilon(T)] - 1 \quad, \qquad (13)$$

where $\varepsilon(T) = \int_0^T \beta dT$ is the volume thermal expansion. In obtaining eq. (13), we assume that the Gruneisen constants $\gamma_{qj}$ do not depend on temperature.

A calculation of the implicit shift, as is evident from eq. (13), requires a calculation of (i) the Gruneisen constants and (ii) the volume thermal expansion.

Expressions for the calculation of thermal expansion, especially for cubic crystals exist already in literature[10,11]. The expression for the lowest-order contribution to the thermal expansion valid for our molecular crystal can be obtained by minimizing the free energy of the crystal, i.e., eq. (6). Expressing the strained lattice volume $V = V(T)$ in terms of unstrained volume $V_0 = V(0)$ and the volume strain parameter $\varepsilon$, i.e.

$$V = V_0(1+\varepsilon), \qquad (14)$$

one can write the static lattice contribution to the free energy as

$$\Phi(V) \approx \Phi(V_0) + \frac{1}{2}\varepsilon^2 V_0^2 \left(\frac{\partial^2 \Phi}{\partial V^2}\right)_{V_0}, \qquad (15)$$

For a structure which gives minimum of potential energy, $\left(\frac{\partial \Phi}{\partial V}\right)_{V_0} = 0$ and eq. (15) can be written as

$$\Phi(V) = \Phi(V_0) + \frac{1}{2}\varepsilon^2 V_0 B, \qquad (16)$$

where, $B$ is the Bulk Modulus and is by definition given by

$$B = V_o \left(\frac{\partial^2 \Phi}{\partial V^2}\right)_{V_o} \qquad (17)$$

Therefore, the condition for the minimum of free energy,

$$\frac{\partial F_{qh}}{\partial \varepsilon} = \varepsilon V_0 B - \frac{1}{2}\sum_{qj} \gamma_{qj} \hbar \omega_{qj} \coth \frac{\hbar \omega_{qj}}{2kT} = 0,$$

gives the lowest order volume thermal expansion as

$$\varepsilon = \frac{1}{2V_o B} \sum_{qj} \gamma_{qj} \hbar \omega_{qj} \coth \frac{\hbar \omega_{qj}}{2kT} \qquad (18)$$

The volume expansion coefficient $\beta$ is then given as

$$\beta = \frac{k}{V_o B} \sum_{qj} \gamma_{qj} \left(\frac{\hbar \omega_{qj}}{kT}\right)^2 \frac{e^{\hbar \omega_{qj}/kT}}{\left(e^{\hbar \omega_{qj}/kT} - 1\right)^2} \tag{19}$$

We would again emphasize that energy and volume expressed here are consistent and are expressed as per atom.

Futher, a knowledge of phonon frequencies enables us to calculate the heat capacity at constant volume, $C_v$.

$$C_v = \left(\frac{\partial E}{\partial T}\right)_v \tag{20}$$

where, internal energy $E$, by definition, is given by

$$E = F - T\left(\frac{\partial F}{\partial T}\right)_v = \sum_{qj} \left\{\Phi(V) + \frac{\hbar \omega_{qj}}{2} + \frac{\hbar \omega_{qj}}{\exp[\hbar \omega_{qj}/kT] - 1}\right\} \tag{21}$$

and thus,

$$C_v = k \sum_{qj} \left(\frac{\hbar \omega_{qj}}{2kT}\right)^2 \bigg/ \sinh^2 \frac{\hbar \omega_{qj}}{2kT} \tag{22}$$

**3.    Numerical Calculations**

Using approximations and procedure as described above, harmonic phonon frequencies have been calculated for various phonon branches for carbon nanotubes of different diameters. The results have been tabulated in table II and phonon dispersion curves for (10,10) tube in two symmetry directions is shown in fig. 2.

The direct lattice vectors of 2-D unit cell are given by

$$\mathbf{a} \equiv (a,0), \quad \mathbf{b} \equiv (\frac{a}{2}, \frac{a\sqrt{3}}{2})$$

Now, a grid of q-vectors (wave vectors) in the Brlliouin zone can be generated by defining

$$\mathbf{q} = l_1 \mathbf{a}^* + l_2 \mathbf{b}^*$$

Where, $l_1$ and $l_2$ are integers and $\mathbf{a}^*$ and $\mathbf{b}^*$ are reciprocal lattice vectors. The reciprocal lattice vectors $\mathbf{a}^*$ and $\mathbf{b}^*$ are defined in the Cartesian axes as

$$\mathbf{a}^* \equiv (\frac{2\pi}{a}, -\frac{2\pi}{\sqrt{3}a}), \text{ and}$$

$$\mathbf{b}^* \equiv \left(0, \frac{4\pi}{\sqrt{3}a}\right)$$

$\mathbf{a}^*$ is perpendicular to $\mathbf{b}$ and $\mathbf{b}^*$ is perpendicular to $\mathbf{a}$.

**Table II:** Phonon frequencies calculated for the SWNT crystals at $\mathbf{q} = 0$ and $\mathbf{q} = (0.5, 0)$

| Br.No. | Character | q-vector | $\omega_{qj}$ (THz) | | |
|---|---|---|---|---|---|
| | | | (10,10) | (17,0) | (12,6) |
| 1 | Translational (longitudinal) | 0.0 | 0.0 | 0.0 | 0.0 |
| | | 0.5 | 1.61 | 1.58 | 1.64 |
| 2 | Translational (transverse) | 0.0 | 0.0 | 0.0 | 0.0 |
| | | 0.5 | 0.892 | 0.94 | 0.93 |
| 3 | Librational (along a*-axis) | 0.0 | 0.052 | 0.05 | 0.053 |
| | | 0.5 | 0.852 | 0.95 | 0.95 |
| 4 | Librational (along b*-axis) | 0.0 | 0.052 | 0.049 | 0.051 |
| | | 0.5 | 1.535 | 1.57 | 1.62 |

**Table III**: Average Gruneisen parameter and Bulk Modulus for clusters of different diameter SWNTs

| Tube | Radius (in Å) | Lattice constant (in Å) | $\frac{1}{4n}\sum_{q=1}^{n}\sum_{j=1}^{4}\gamma_{qj}$ | Bulk modulus (GPa) | |
|---|---|---|---|---|---|
| | | | | ours | Other[12] |
| (5,5) | 3.46 | 9.885 | 11.04 | 36.2 | 36.16 |
| (10,10) | 6.78 | 16.672 | 18.73 | 51.6 | 50.91 |

| (15,15) | 10.32 | 23.44 | 26.27 | 64.09 | 64.74 |

We present here, in Table III, the results of the average Gruneisen parameter summation for tubes having different diameters and it is found that this value strongly depends on diameter and this implies the larger the molecule, the larger will be the value.

From eq. (4), one can easily obtain the P-V curve, as shown in Fig. 2, which enables one to evaluate the bulk modulus given by eq.(17), given in Table III, the values of bulk modulus that we obtain for different diameter tubes are in excellent agreement with those given by Popov et al. [12] for rigid bundles.

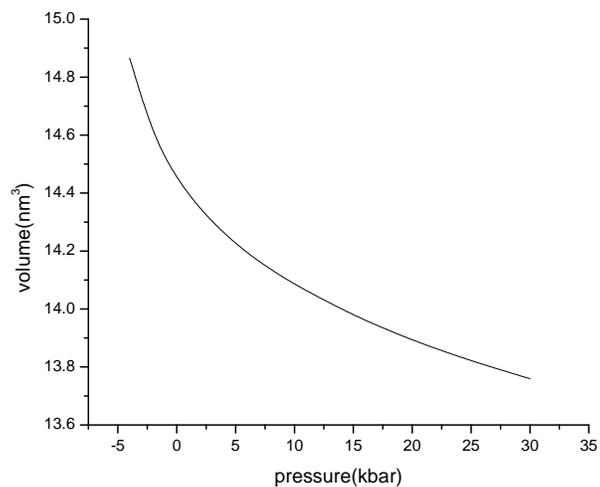

**Fig. 2**: P-V curve for (10,10) nanorope

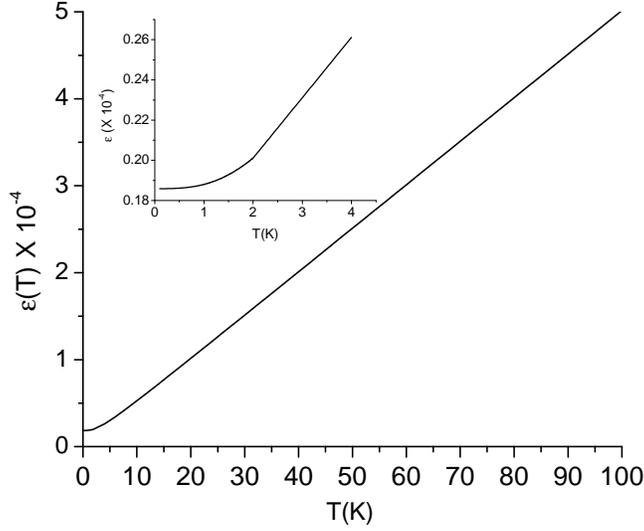

**Fig.3**: Volume thermal expansion $\varepsilon(T)$ as a function of temperature. Inset shows low-temperature behaviour of thermal expansion.

Using the phonon frequencies, the Gruneisen constants and the bulk modulus, the sum over the wave-vector etc. over the first Brillouin zone can be easily carried out. This leads us to compute easily the volume thermal expansion and its coefficient, as given by eqs. (18) and (19), as a function of temperature. In our model, since the tubes are considered to be rigid, the hydrostatic pressure does not alter their length and only lattice parameter is changed. Therefore, the calculation of $\varepsilon$ and $\beta$ will then give only the expansion coefficient related to lattice parameter. In the calculation of $\varepsilon$ and $\beta$, we took the length of the tube equal to its lattice parameter. The values of thermal expansion coefficient for tubes having different diameters have been calculated and tabulated in Table IV. Bulk modulus comes out to be independent of length of nanotubes in a nanorope. However, $\varepsilon$ and $\beta$ decrease by increasing the length of CNTs.

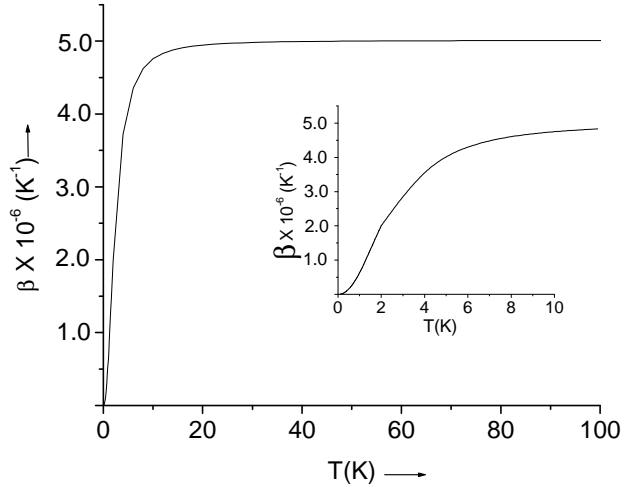

**Fig. 4**: Volume thermal expansion coefficient $\beta$ as a function of temperature. Inset shows low-temperature behaviour of thermal expansion coefficient.

Our calculated value of volume thermal expansion coefficient $\beta$ for (10,10) nanotube bundle near 300 K is around $0.5 \times 10^{-5}$/K, which is close to the value of $0.75 \times 10^{-5}$/K that was determined by Maniwa et.al.[13] for the triangular lattice constant by means of x-ray scattering. There is no other calculation or measurement available in the literature. The calculated results for lowest order $\varepsilon$ and $\beta$ for a bundle consisting of (10,10) tubes are given in Figs. 3 and 4 respectively.

**Table IV:** Thermal expansion coefficient at room temperature

| Tube | $\beta$ (X $10^{-5}$ K$^{-1}$) | $\beta$ (X $10^{-5}$ K$^{-1}$) [123 |
|---|---|---|
| (5,5) | 2.27 | |
| (10,10) | 0.50 | 0.75 |
| (15,15) | 0.199 | |

Now, it is easy to calculate the implicit shift in the phonon frequencies, once the essential ingredients have been calculated as described. This implicit frequency shift is temperature dependent as $\varepsilon$ is linearly dependent on temperature at high temperatures and the gradient of temperature dependence is governed by the Gruneisen parameter for that mode. We plot in Fig. 5 this temperature dependence for two librational modes.

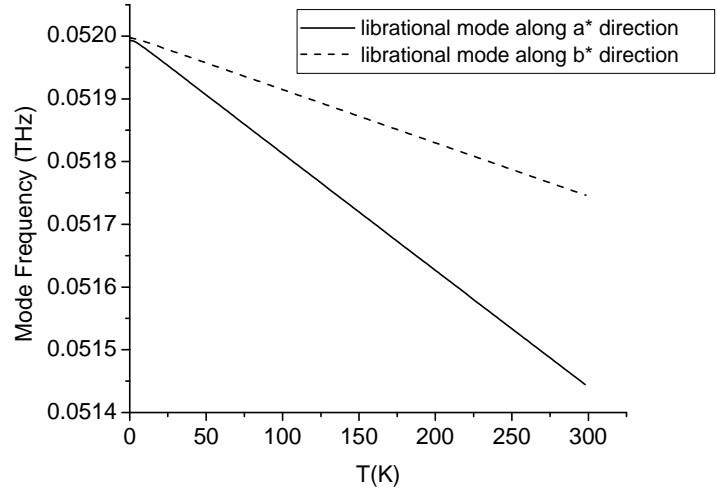

**Fig. 5**: Temperature dependence of the calculated librational modes.

Heat Capacity, given by eq. (23), has also been calculated and plotted as a function of temperature in fig.6.

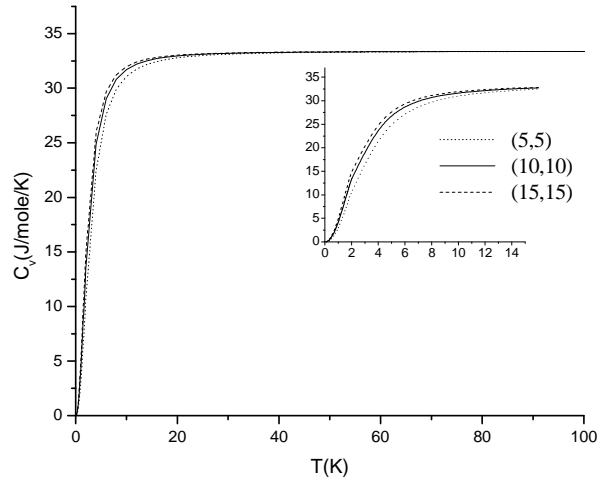

**Fig. 6**: Temperature dependence of calculated heat capacity. Inset shows low-temperature behaviour of heat capacity.

## 4. Discussions and Conclusions

In this paper, we have attempted to present the various temperature dependent properties of nanotube bundles using a simple atom-atom potential for intermolecular interactions in the solid. Using this simple potential, the harmonic phonons have been calculated and an earlier calculation [14] also indicates that these results fall in the same range as obtained here, (5cm$^{-1}$ to 60 cm$^{-1}$). For the case of 2-D system, the motion along "c" axis, which is taken to coincide with the z-Cartesian axis is quenched. This reduces the vibrational modes to 2 per molecule. Minimization of energy at various mutual orientations of CNTs in a rope reveal that the differences in energy due to orientational repositioning of the nanotubes was very small [4]; consequently the librational mode along 'c' axis was very low. A libration along, say, a$^*$ direction involves the libration of the long cylinder along a direction perpendicular to the long axis. In a realistic bunch of long nanotubes, librations along directions perpendicular to long axis would have to be very low in amplitude.

In table III, we have presented the Gruneisen parameter averaged over all phonon modes for different diameter nanotubes. We have not been able to get the values of Gruneisen parameter for carbon nanotubes anywhere in the literature for comparison. From our results, we find that this parameter is system dependent. The larger the molecule, the larger is its value. However, the bulk modulus that we have calculated corresponds nicely with the bulk modulus given by Popov et al. [12] who, for a crystal of rigid tubes with VdW interactions, showed that the moduli can be fitted with a polynomial of second degree- $K(GPa) = 20.66 + 4.828R - .054R^2$ where $R$ is in Å.

Further, in fig.5, we have plotted the temperature dependence of some modes by calculating the implicit shifts in the phonon frequencies. A quantitative estimate of the frequency shift would have to include explicit anharmonicity. The implicit contribution is dominant and negative, whereas the explicit contribution is normally positive [5], so that the net effect is likely to result in reduced magnitude of shifts. As far as our knowledge is concerned, there is no measurement for phonon shifts of rigid CNT bundles available in the literature and it would be useful to get experimental information about these shifts. However, some measurements are available for SWNTs and MWNTs [13, 15-18].

The calculated volume thermal expansion coefficient, where only lattice parameter is the variable and there is no alteration in the length, is available from fig.4. Its value for the bundles consisting of about 14 Å diameter tubes ($0.5\times 10^{-5}$/K around room temperature), with length equal to that of lattice parameter, compares reasonably well with that given by x-ray scattering experiments [13]($\sim 0.75\times 10^{-5}$/K). From our calculations, we find that volume thermal expansion and its coefficient decreases with increase in length of CNTs in a rope.

We have also calculated the heat entropy for the nanotube bundles and we find that it approaches to the classical value at high temperatures. We have not taken into account any intramolecular interactions and calculated the specific heat of the bundles. However, all other authors consider the tubes to be flexible within the bundles and then calculated the heat entropy [19-22].

In conclusion, we have studied the thermodynamical properties of carbon nanotube materials using only weak VdW potential between the tubes, assuming the tubes to be rigid. Dominant temperature dependent features of these rigid carbon nanotube bundles have been studied and the values of thermal expansion coefficient and bulk moduli are found to be consistent with the literature. The modifications that incorporate the effect of flexibility of tubes need to be done which will motivate more experimental as well as theoretical work. However such deformations are prominent for larger diameter tubes or when an external strain is applied perpendicular to the long axis of the tubes. Even an isolated tube assumes a flattened or collapsed structure whose extent depends largely on the diameter of the tube [23]. Such pressure induced phase transitions have already been studied in Ref [6]. From that model, we found that about 34Å diameter tubes show faceting even at ambient conditions; however Tersoff [24] found this for 25Å diameter tubes and Lopez et al. [25] observed this faceting for 17Å tubes. Further, it is harder to distort a multiwalled nanotube (MWNT) than a single walled nanotube (SWNT) [26].

Thus, the work presented in this paper on thermodynamics of rigid ropes is suitable for smaller diameter SWNT crystals as well as MWNT where the rigid molecule model works well. The implicit phonon shifts have been calculated using the thermal expansion of a bunch of tubes. Since the VdW forces between the tubes in a rope are much weaker as well as much more anharmonic as compared to strong intramolecular forces, the dominant contribution, while calculating the thermal expansion of a rope, is from the change in intertube spacing and not from the deformations of the individual tubes. Due to this reason of different scales of magnitude of inter- and intra-molecular interactions, the external modes are well separated from the internal modes. Therefore the shifts in frequencies of external modes would also be reasonably independent of the internal modes, indicating thereby that the inclusion of the flexibility of the tubes would not have any significant impact on the phonon shifts presented here. However the thermodynamics of flexible carbon nanotubes is equally interesting and work on this is in progress and will be published separately. As far the results of extensive quantities like cohesive energy etc. it can be scaled up as per the lenth of the ropes.

Acknowledgements:

VKJ wishes to acknowledge Council for Scientific and Industrial Research, New Delhi for support as Emeritus Scientist. He also acknowledges the support from Alexander von Humboldt Foundation, Germany for a revisit to complete this work.

**<u>Figure Captions</u>**

**Fig.1**: Cross-section through a bundle of carbon nanotube.

**Fig. 2**: P-V curve for (10,10) nanorope

**Fig.3**: Volume thermal expansion $\varepsilon(T)$ as a function of temperature. Inset shows low-temperature behaviour of thermal expansion.

**Fig. 4**: Volume thermal expansion coefficient $\beta$ as a function of temperature. Inset shows low-temperature behaviour of thermal expansion coefficient.

**Fig. 5**: Temperature dependence of the calculated librational modes.

**Fig. 6**: Temperature dependence of calculated heat capacity. Inset shows low-temperature behaviour of heat capacity.

**Table I:** Atom-atom potential parameters (Kitaigorodski [9])

| A=358 kcal/mole-$A^6$ | B=42000 kcal/mole | $\alpha$=3.58$A^{-1}$ |

**Table II:** Phonon frequencies calculated for the SWNT crystals at **q** = 0 and **q** = (0.5,0)

| Br.No. | Character | q-vector | $\omega_{qj}$ (THz) | | |
|---|---|---|---|---|---|
| | | | (10,10) | (17,0) | (12,6) |
| 1 | Translational (longitudinal) | 0.0 | 0.0 | 0.0 | 0.0 |
| | | 0.5 | 1.61 | 1.58 | 1.64 |
| 2 | Translational (transverse) | 0.0 | 0.0 | 0.0 | 0.0 |
| | | 0.5 | 0.892 | 0.94 | 0.93 |
| 3 | Librational (along a*-axis) | 0.0 | 0.052 | 0.05 | 0.053 |
| | | 0.5 | 0.852 | 0.95 | 0.95 |
| 4 | Librational (along b*-axis) | 0.0 | 0.052 | 0.049 | 0.051 |
| | | 0.5 | 1.535 | 1.57 | 1.62 |

**Table III**: Average Gruneisen parameter and Bulk Modulus for clusters of different diameter SWNTs

| Tube | Radius (in Å) | Lattice constant (in Å) | $\frac{1}{4n}\sum_{q=1}^{n}\sum_{j=1}^{4}\gamma_{qj}$ | Bulk modulus (GPa) | |
|---|---|---|---|---|---|
| | | | | ours | Other[12] |
| (5,5) | 3.46 | 9.885 | 11.04 | 36.2 | 36.16 |
| (10,10) | 6.78 | 16.672 | 18.73 | 51.6 | 50.91 |
| (15,15) | 10.32 | 23.44 | 26.27 | 64.09 | 64.74 |

**Table IV:** Thermal expansion coefficient at room temperature

| Tube | $\beta$ (X $10^{-5}$ $K^{-1}$) | $\beta$ (X $10^{-5}$ $K^{-1}$) [12] |
|---|---|---|
| (5,5) | 2.27 | |
| (10,10) | 0.50 | 0.75 |
| (15,15) | 0.199 | |

**Fig. 1**

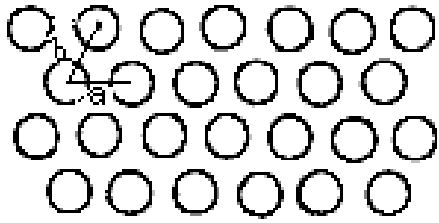

**Fig. 2**

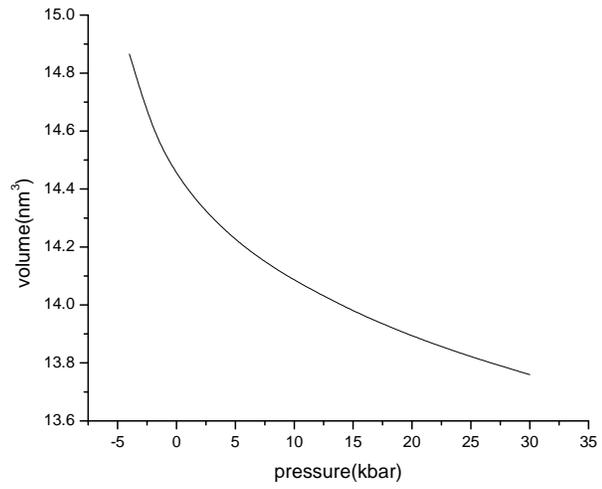

**Fig. 3**

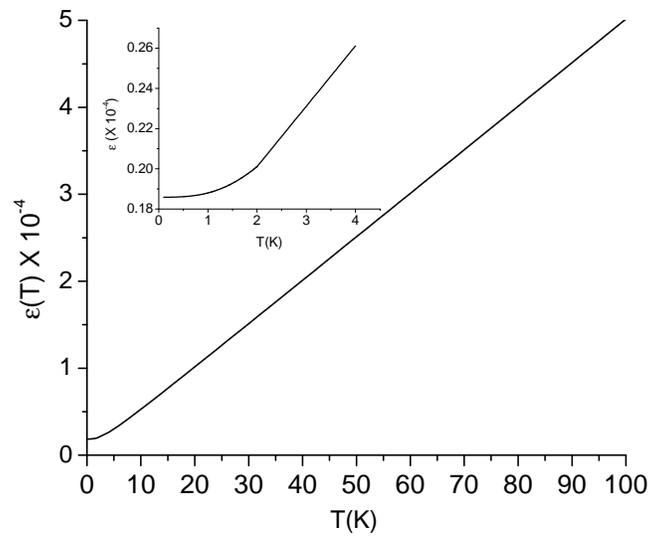

**Fig. 4**

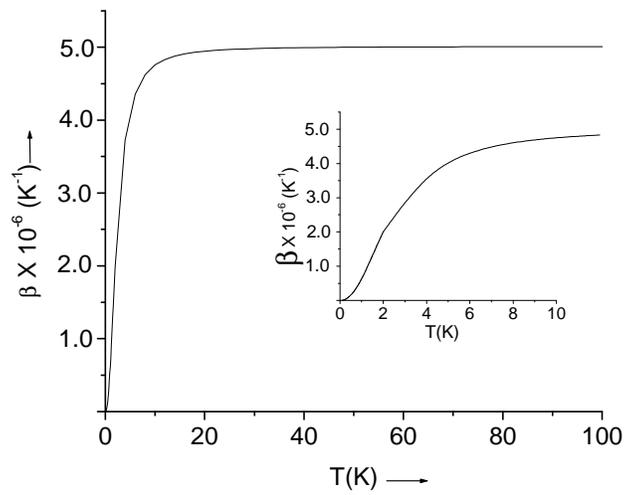

**Fig. 5**

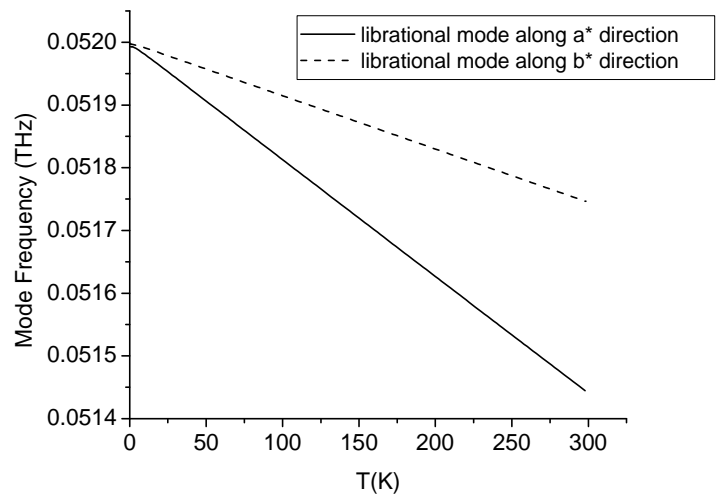

**Fig. 6**

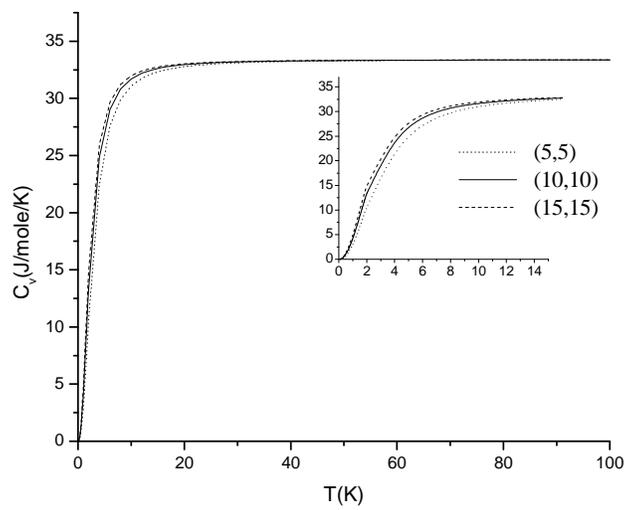